%% file: llncs.tex
\documentclass{llncs}

\usepackage{amsmath, dsfont, braket,todonotes,caption}
\usepackage{tikz} 
\usepackage{wrapfig}
\usepackage{listings}
\usepackage{amsmath}
\usepackage{ifthen}
\usepackage{hyperref}
\usepackage{pgfplots}
\usepackage{fontawesome}
\usepackage{verbatim}
\usepackage{lscape}
\usepackage{microtype}
\usepackage[numbers,sort]{natbib}
\makeatletter
\renewcommand\bibsection%
{
  \section*{\refname
    \@mkboth{\MakeUppercase{\refname}}{\MakeUppercase{\refname}}}
}
\makeatother
\usepackage{booktabs}

\definecolor{pantone280}{RGB}{100,100,100}

\definecolor{hks33}{RGB}{155,10,125}
\definecolor{hks37k}{RGB}{086,035,129}
\definecolor{hks12k}{RGB}{228,049,023}
\definecolor{pantone382}{RGB}{153,194,033}
\definecolor{pantone292}{RGB}{106,172,218}
\definecolor{hks63k}{RGB}{057,132,046}
\definecolor{hks6k}{RGB}{242,148,000}
\definecolor{pantone315}{RGB}{000,103,124}
\definecolor{pantone280s}{cmyk}{100,080,000,0.15}
\definecolor{pantone280s}{RGB}{000,061,143}

\let\origthelstnumber\thelstnumber
\makeatletter

\def\lst@PlaceNumber{\ifnum\value{lstnumber}=0\else
	\llap{\normalfont\lst@numberstyle{\thelstnumber}\kern\lst@numbersep}\fi}

\newcommand*\Suppressnumber{%
	\lst@AddToHook{OnNewLine}{%
		\let\thelstnumber\relax%
		\advance\c@lstnumber-\@ne\relax%
	}%
}

\newcommand*\Reactivatenumber[1]{%
	\setcounter{lstnumber}{\numexpr#1-1\relax}
	\lst@AddToHook{OnNewLine}{%
		\let\thelstnumber\origthelstnumber%
		\refstepcounter{lstnumber}
	}%
}

\usetikzlibrary{positioning, shapes,patterns,shadows.blur, arrows,decorations.text,arrows,automata,shadows,patterns,chains,bending,arrows.meta}
\tikzset{%
	initial text={},
	nonconsideredstate/.style={draw=black,fill=pantone280},
	smallState/.style={minimum size=0.55cm},
	bigState/.style={minimum size=1cm},
	lightText/.style={color=pantone280},
	consideredState/.style={draw=black!50,fill=pantone280!30},
	notConsideredState/.style={color=pantone280!20,left color=pantone280!10, right color=pantone280!20,shading angle=155,text=pantone280!20,minimum size=0.55cm,font=\tiny, blur shadow={shadow blur steps=5,shadow yshift=-0.1, shadow xshift=-0.1}},
	dataStyle/.style={draw=black,fill=white,text=black,minimum size=0.55cm,font=\tiny,anchor=north,rounded corners=2},
	datalabel/.style={draw=black,fill=black!5,text=black,font=\tiny,rectangle,rounded corners=3},
	data1/.style={dataStyle,draw,text centered,minimum width=1.5cm},
	data2/.style={dataStyle,draw,rectangle split,rectangle split parts=2,text centered,minimum width=1.5cm},
	data2+/.style={data2,rectangle split every empty part={},
		rectangle split empty part width=\widthof{#1},rectangle split empty part height=\heightof{#1},rectangle split empty part depth=\depthof{#1},},
	data3/.style={dataStyle,draw,rectangle split,rectangle split parts=3,text centered,minimum width=1.5cm},
	data3+/.style={data3,rectangle split every empty part={},
		rectangle split empty part width=\widthof{#1},rectangle split empty part height=\heightof{#1},rectangle split empty part depth=\depthof{#1},},
}
\newcommand{\dotequal}{\,\dot{=}\,}
\newcommand\pto{\mathrel{\ooalign{\hfil$\mapstochar$\hfil\cr$\to$\cr}}}
\renewcommand{\epsilon}{\varepsilon}
\newcommand{\variableVars}{\mathsf{K}}
\newcommand{\constsVars}{\mathsf{C}}
\newcommand{\stateVars}{\mathsf{S}}

\newcommand{\wordEncoding}{\mathsf{word}}
\newcommand{\wordsMatch}{\mathsf{wm}}
\newcommand{\onehot}{\mathsf{OH}}
\newcommand{\mddVar}{\mathsf{M}}
\newcommand{\coefficients}{\mathsf{Co}_{u,v}}
\newcommand{\filledWord}[1]{#1_\xi}
\newcommand{\filledSubs}[1]{S_\xi\left(#1\right)}
\newcommand{\filledLambdaPattern}{\Xi_{\xi_\lambda}}
\newcommand{\congrel}{\overset{\filledWord{S}}{\sim}}

\newcommand{\subs}[2]{\left[\frac{#1}{#2}\right]}

\newcommand{\fakeMathCenter}[1]{\bgroup\par\centering#1\par\egroup\noindent}

\renewcommand{\epsilon}{\varepsilon}

\pgfplotscreateplotcyclelist{my black white}{%
	solid, every mark/.append style={solid, fill=pantone280}, mark=*\\%
	dotted, every mark/.append style={solid,fill=pantone280!80}, mark=square*\\%
	densely dotted, every mark/.append style={solid, fill=pantone280!50}, mark=otimes*\\%
	loosely dashed, every mark/.append style={solid, fill=pantone280!20}, mark=triangle*\\%
	dashed, every mark/.append style={solid, fill=gray},mark=diamond*\\%
	loosely dotted, every mark/.append style={solid, fill=gray},mark=x\\%
	densely dashed, every mark/.append style={solid, fill=gray},mark=square*\\%
	dashdotted, every mark/.append style={solid, fill=gray},mark=otimes*\\%
	dashdotdotted, every mark/.append style={solid},mark=star\\%
	densely dashdotted,every mark/.append style={solid, fill=gray},mark=diamond*\\%
}

\DeclareMathOperator{\subsat}{\models}
\DeclareMathOperator{\var}{var}

\newcommand{\woorpje}{\textsc{Woorpje}}%

\begin{document}

\title{On Solving Word Equations Using SAT}
\author{Joel D. Day\inst{1}, Thorsten~Ehlers\inst{2},  Mitja~Kulczynski\inst{3}, Florin~Manea\inst{3}, Dirk~Nowotka\inst{3} and Danny~B\o gsted~Poulsen\inst{3}}
\authorrunning{J. Day, T. Ehlers,  M. Kulczynski, F. Manea, D. Nowotka and D. Poulsen}

\institute{
	Department of Computer Science, Loughborough University,\\
	\and
	German Aerospace Center (DLR), Helmholtz Association,\\
	\and
	Department of Computer Science, Kiel University,\\
    }
\maketitle

\begin{abstract}
We present \textsc{Woorpje}, a string solver for bounded word equations (i.e., equations where the length of each variable is upper bounded by a given integer). Our algorithm works by reformulating the satisfiability of bounded word equations as a reachability problem for nondeterministic finite automata, and then carefully encoding this as a propositional satisfiability problem, which we then solve using the well-known Glucose SAT-solver. This approach has the advantage of allowing for the natural inclusion of additional linear length constraints.
Our solver obtains reliable and competitive results and, remarkably, discovered several cases where state-of-the-art solvers exhibit a faulty behaviour.

\end{abstract}

\section{Introduction}\label{sec:Introduction}
\input{atvaintro}

\section{Preliminaries}\label{sec:Preliminaries}
\input{preliminaries}

\section{Word equation solving}\label{sec:wes}
\input{wordequationsolving}

\section{Experiments}\label{sec:compare}
\input{experiments}

\section{Conclusion} \label{sec:conclusion}
\input{conclusion}
	

\bibliographystyle{splncsnat}
\bibliography{main}

\newpage
\appendix

\section{Artifacts for this paper}\label{sub:artifcats}
You can download our artifacts at \url{http://informatik.uni-kiel.de/~mku/woorpje/rp19.tar.gz}. This package contains:
\begin{enumerate}
	\item A \texttt{README.txt} including a guide to use our artifacts.
	\item Download script for all used tools
	\item Benchmark suite
	\item \woorpje{} source code
\end{enumerate}

\section{Preprocessing algorithms}\label{sub:preprocessing}
\subsection{Prefix reducer}
	For a word $w \in \Xi^*$ we define the prefix of length $i \in [|w|]$ by $w[i:] = w[1] \dots w[i]$. This algorithm is used analogously for suffices.
	\begin{lstlisting}[mathescape=true,caption={Prefix reducer},label=alg:prefixreducer]
|\Suppressnumber|$\mathsf{Input:}$ $u \dotequal v$ # input word equation
$\mathsf{Output:}$ $u' \dotequal v'$ # reduced word equation|\hrule||\Reactivatenumber{2}|
commonPrefixIndex = 0
for $i \in \min\Set{|v|,|u|}$
	if $v[i] \neq u[i]$
		commonPrefixIndex = i
		break
return $u[commonPrefixIndex:] \dotequal v[commonPrefixIndex:]$
\end{lstlisting}
	\begin{example}
		Consider the word equation $aaX \dotequal aabY$ where capital letters are variables and lower case letters are constants. The above algorithm reduces the equation to $X \dotequal bY$.
	\end{example}
\subsubsection{Prefix mismatch}
This algorithm is used analogously for suffices.
\begin{lstlisting}[mathescape=true,caption={Prefix mismatch},label=alg:MDD]
|\Suppressnumber|$\mathsf{Input:}$ $u \dotequal v$ # input word equation
$\mathsf{Output:}$ $x \in \Set{\texttt{UNSAT},\texttt{UNKNOWN}}$|\hrule||\Reactivatenumber{2}|
for $i \in \min\Set{|v|,|u|}$
	if $v[i],u[i] \in \Gamma$:
		break
	if $v[i] \neq u[i] \land v[i],u[i] \in \Sigma$:
		return UNSAT
return UNKNOWN
\end{lstlisting}
\begin{example}
	Consider the word equation $abX \dotequal aabY$ where capital letters are variables and lower case letters are constants. The above algorithm returns \texttt{UNSAT} due to the mismatch of the second characters.
\end{example}

\subsection{Constant sequence mismatch}
Within the following algorithm we use the set of factors of a word $w \in \Xi^*$ which is defined by $\mathsf{factor}(w) = \Set{v \in \Xi | \exists\;u_1,u_2 \in \Xi^* : w = u_1 v u_2 }$. For a given equation $u \dotequal v$ this algorithm assumes $u \in \Sigma^*$ and $v \in \Xi^*$.
\begin{lstlisting}[mathescape=true,caption={Constant sequence mismatch},label=alg:MDD]
|\Suppressnumber|$\mathsf{Input:}$ $u \dotequal v$ # input word equation
$\mathsf{Output:}$ $x \in \Set{\texttt{UNSAT},\texttt{UNKNOWN}}$|\hrule||\Reactivatenumber{2}|
for $w \in \Set{x \in \Sigma^* | x \in \mathsf{factor}(v) \land \forall\,y \in \mathsf{factor}(v): y \in \mathsf{factor}(x) \Rightarrow |x| \geq |y|}$
	if $w \notin \mathsf{factor}(u)$
		return UNSAT
return UNKNOWN
\end{lstlisting}
  
\begin{example}
	Consider the word equation $ababab \dotequal XaabY$ where capital letters are variables and lower case letters are constants. The above algorithm returns \texttt{UNSAT} because $aab \notin \mathsf{factor}(ababab)$.
\end{example}

\subsection{Parikh matrix mismatch}
This algorithm is used analogously for suffices. It uses the Parikh matrix which for a word $w \in \Xi^*$ is defined for each $a \in \Xi$ and $i \in [|w|]$ by $M_w[i][a] = |w[:i]|_a$.
\begin{lstlisting}[mathescape=true,caption={Parikh matrix mismatch},label=alg:MDD]
|\Suppressnumber|$\mathsf{Input:}$ $u \dotequal v$ # input word equation
$\mathsf{Output:}$ $x \in \Set{\texttt{UNSAT},\texttt{UNKNOWN}}$|\hrule||\Reactivatenumber{2}|
for $i \in \min\Set{|v|,|u|}$
	variablesAlign = true
	terminalsAlign = true
	for $a \in \Xi$
		if $M_v[i][a] \neq M_v[i][a]$:
			if $a \in \Gamma$
				variablesAlign = false
			else
				terminalsAlign = false
	if variablesAlign and not terminalsAlign
		return UNSAT
return UNKNOWN
\end{lstlisting}
\begin{example}
	Consider the word equation $aX \dotequal Xb$ where capital letters are variables and lower case letters are constants. The above algorithm returns \texttt{UNSAT} due to a Parikh matrix mismatch within index 2 regarding character $a$ (and $b$). The matrices look as following:
		$$\bordermatrix{ 
			&{\color{black!50} \text{\tiny $1$} } & 
			{\color{black!50} \text{\tiny $2$} }  \cr
			{\color{black!50} \text{\tiny$a$}} 	& 1 & 1 \cr
			{\color{black!50} \text{\tiny$b$}}	& 0 & 0 \cr 
		    {\color{black!50} \text{\tiny$X$}}	& 0 & 1 \cr } \hspace*{1cm}\text{and} \hspace*{1cm}
	    	\bordermatrix{ 
		    &{\color{black!50} \text{\tiny $1$} } & 
		    {\color{black!50} \text{\tiny $2$} }  \cr
		    {\color{black!50} \text{\tiny$a$}} 	& 0 & 0 \cr
		    {\color{black!50} \text{\tiny$b$}}	& 0 & 1 \cr 
		    {\color{black!50} \text{\tiny$X$}}	& 1 & 1 \cr }$$	
\end{example}

\subsection{Substitution reasoning}
For a system of word equations $E$ this algorithm assumes that all equations $u \dotequal v \in E$ are minimized with our algorithm given in \autoref{alg:prefixreducer}.
\begin{lstlisting}[mathescape=true,caption={Substitution reasoning},label=alg:MDD]
|\Suppressnumber|$\mathsf{Input:}$ $E$ # a System of word equations
$\mathsf{Output:}$ $(E,x)$ # $E $ - a system of equations, $x \in \Set{\text{SAT},\texttt{UNSAT},\texttt{UNKNOWN}}$|\hrule||\Reactivatenumber{2}|
E' = $\Set{u \dotequal v | u \in \Gamma \land v \in \Sigma^* \land \left(u \dotequal v \in E \lor v \dotequal u \in E \right)}$

if E' = $\emptyset$
	return ($\emptyset$,UNKNOWN)
	
for $u \dotequal v \in E$
	for $A \dotequal w \in E'$
		$v'$ = $S(u)[A \mapsto w]$
		$u'$ = $S(v)[A \mapsto w]$
		if $u \neq u' \lor v \neq v'$
			if v $\neq$ u
				if applyOtherUNSATalgorithms($u \dotequal v$) = UNSAT
					return ($\emptyset$,UNSAT)
				v = v'
				u = u'
			else
				$E$ = $E \setminus u \dotequal v$
				
if E = $\emptyset$
	return (E',SAT)
return(E',UNKNOWN)
\end{lstlisting}
\begin{example}
	Consider the word equations $X \dotequal aab$,$Y \dotequal a$ and $aX = Yaab$ where capital letters are variables and lower case letters are constants. The above algorithm returns \texttt{SAT} and the set of equations $X \dotequal aab$,$Y \dotequal a$, which gives us the substitutions $S(X) = aab$ and $S(Y) = a$.
\end{example}

\section{More detailed benchmark results}
\begin{figure}[t]

\resizebox{.45\textwidth}{!}{
}

	\caption{Cactus plots for all Tracks\label{fig:cactusAll}}
\end{figure}

\begin{landscape}
\begin{table}[t]
	\centering
	\resizebox{20cm}{!}{
		\begin{tabular}{*{41}{c}}
			& \multicolumn{5}{c}{\textsc{Woorpje}} & & \multicolumn{5}{c}{\textsc{CVC4}}& & \multicolumn{5}{c}{\textsc{z3Str3}}& & \multicolumn{5}{c}{\textsc{z3Seq}} & &\multicolumn{5}{c}{\textsc{Norn}} & &\multicolumn{5}{c}{\textsc{Sloth}}\\
			\cmidrule(l){2-6}\cmidrule(l){8-12}\cmidrule(l){14-18}\cmidrule(l){20-24}\cmidrule(l){26-30}\cmidrule(l){32-36}
			& \tiny I & \tiny II & \tiny III & \tiny IV &\tiny  V & & \tiny I & \tiny II & \tiny III & \tiny IV &\tiny  V && \tiny I & \tiny II & \tiny III & \tiny IV &\tiny  V & &\tiny I & \tiny II & \tiny III & \tiny IV &\tiny  V & &\tiny I & \tiny II & \tiny III & \tiny IV &\tiny  V & &\tiny I & \tiny II & \tiny III & \tiny IV &\tiny  V   \\
\end{tabular}}
	\caption{Benchmark results}
	\label{tab:resultsApp}
\end{table}
\end{landscape}

	\autoref{tab:resultsApp} is read as follows: \texttt{sat} is the count of instances classified as satisfiable, where \texttt{unsat} marks the unsatisfiable cases.
For instances marked as \texttt{unknown} the solver returned no answer but terminated before the timeout was reached, where in \texttt{timeout} cases the solver was killed after 30 seconds. The row  \texttt{error} lists the count of incorrectly classified instances. The row which states the time lists the overall solving time with and without timeout cases. The produced substitutions were checked regarding their correctness afterwards. The classification of an error was done by ad-hoc case inspection whenever not all solvers agreed on a result.  In the cases one solver produced a valid solution, and others did not, we validated the substitutions manually. For the cases where one solver determined an equation is unsatisfied and all others timed out, we treated the unsat result as correct. This means that we only report errors if a solver reports unsat and we know the equation was satisfiable. 
The table lists only 199 benchmarks on Track I for CVC4, since the program crashed with a null-pointer exception regarding the equation $dbebgddbecfcbbAadeeaecAgebegeecafegebdbagddaadbddcaeeeb\-fabfefabfacdgAgaabgegagf \dotequal dbebgddbeAfcbbAaIegeeAaDegagf$, where lowercase symbols are letters and uppercase letters are variables. This word equation is stored as benchmark 111 in Track I. 

The erroneous behavior of Sloth was detected within the following benchmarks: Track III; Benchmarks 10,11,14,15,16,17,18,21,23,24,4,43 and Track V; 192,53.

\begin{table}[t]
	\centering
		\begin{tabular}{*{7}{c}}
			& \multicolumn{5}{c}{\textsc{Woorpje}}\\
			\cmidrule(l){2-6}
			& \tiny I & \tiny II & \tiny III & \tiny IV &\tiny  V\\
			\cmidrule(l){2-6}
sat& 200 & 5 & 147 & 104 & 162\\
unsat& 0 & 0 & 0 & 0 & 0 \\
\cmidrule(l){2-6}
unknown& 0 & 0 & 6 & 63 & 9\\
timeout& 0 & 4 & 47 & 33 & 29\\
error& 0 & 0 & 0 & 0 & 0\\
Time (s)& \tiny8.10 & \tiny123.62 & \tiny1423.06 & \tiny1162.00 & \tiny885.18\\
\cmidrule(l){2-6}
Time w/o timeouts (s)& \tiny8.10 & \tiny3.62 & \tiny13.06 & \tiny172.00 & \tiny15.18
	\end{tabular}
\caption{Benchmark results without preprocessing}
\label{tab:resultsAppPre}
\end{table}

\autoref{tab:resultsAppPre} shows the results \woorpje{} produces without using preprocessing techniques. It is read in exactly the same way as the previous table. Comparing the times shows that preprocessing not only speeds up the solving time, but also helps to classify a few more positive instances.

\end{document}

%% file: atvaintro.tex
Over the past twenty years, applications of software verification have scaled from
small academic programs to finding errors in the GNU Coreutils~\citep{CadarDE08}. In principle, the employed verification strategies
involve exploring the control-flow-graph of the program, gathering
constraints over program variables and passing these constraints to a
constraint solver. The primary worker of software verification is thus
the constraint solver, and the scalability of software verification
achieved by improving the efficiency of constraint
solvers. The theories supported by constraint solvers are likewise
highly influenced by the needs of software verification tools (e.g. array theory
and bitvector arithmetic). A recent need of software verification
tools is the ability to cope with equations involving string constraints,  i.e. equations over string
variables composing equality between concatenation
of strings and string variables. This need arose from the desire to do
sofware verification of languages with string manipulation as a core part of the language (e.g. JavaScript and
Java)~\cite{saxena2010symbolic,jbmc}. Accomodating for this need, we have seen the advent of
dedicated string solvers as well as constraint solvers implementing
string solving techniques. As an incomplete list we mention HAMPI \cite{kiezun2009},
CVC4~\cite{barrett2011}, Ostrich~\cite{chen2019decision}, Sloth~\cite{holik2017string}, 
Norn~\cite{abdulla2015}, S3P~\cite{trinh2016} and
Z3str3~\cite{berzish2017}.

Although the need for string solving only recently surfaced in the
software verification community, the problem is in fact older and
known as \emph{Word Equations} (a term that we will use from now on). 
The word equation satisfiability problem is to determine whether we can unify the two strings, i.e., transform them 
into two equal strings containing constant letters only, by substituting the
variables consistently by strings of constants. For example, consider 
the 
equation defined by the two strings  $X a bY$ and $a XY b$, denoted 
$X a b Y 
\dotequal a XY b$, with variables $X, Y$ and constants $a$ and $b$. 
It is 
satisfiable because $X$ can be substituted by $a$ and $Y$ by $b$, 
which 
produces the equality $aabb = aabb$.  In fact, substituting $X$ by an 
arbitrary 
amount of $a$'s and $Y$ by an arbitrary amount of $b$'s unifies the 
two sides 
of the equation.

The word equation problem is decidable~\cite{makanin1977} and 
NP-hard. In a series of works, \citet{jez2017,jez2013}
showed that word equations can be solved in non-deterministic linear
space.  It has been shown by
\citet{plandowski1999} that there
exists an upper bound of $2^{2^{O(n^4)}}$ for the smallest solution to
a word equation of length $n$. Having this in mind, a standard method
for solving word equations is known as \emph{filling the
  positions}~\citep{karhumaki2000,PlandowskiR98}. In this method a
length for each of the string variables is non-deterministically
selected. Having a fixed length of the variables reduces the
problem to lining up the positions of the two sides of the equation, and
filling the unknown positions of the variables with characters, making the
two sides equal. 


In this paper we present a new solver for word equation with linear length
constraints, \woorpje.  
In particular, it guesses the maximal length of variables and  encodes a variation of  \emph{filling the positions} method into an
automata-construction, thereby reducing the search for a solution to a
reachability question of this automata. Preliminary experiments with
a pure automata-reachability-based approach revealed however, that
this would not scale for even small word equations.  \woorpje\
therefore encodes the automata into SAT and uses the tool Glucose~\citep{audemard2018glucose} as a backend. Unlike other approaches based on the 
filling the positions method (e.g. \cite{bjorner2009path,saxena2010symbolic}), \woorpje\ does 
not need an exact bound for each variable, but only an upper bound.
Experiments indicate that \woorpje\ is not only reliable but also
competitive to the more mature CVC4 and Z3. Results 
indicate that \woorpje\ is quicker on pure word equations (no linear
length constraints), and that CVC4 and Z3 mainly have an edge  on word
equations with linear constraints. This may be due to our naive solution for
solving linear length constraints.


%% file: preliminaries.tex
Let $\mathds{N}$ be the set of natural numbers, let  $[n]$ be the set
$\Set{0,1,2,\ldots,n-1}$ and $[n]_0$ the set $[n] \setminus
\Set{0}$. For  a finite set  $\Delta$ of symbols,  we let $\Delta^*$ be the set 
of all words over $\Delta$ and~$\varepsilon$ be the empty
word. For an alphabet $\Delta $ and  $a \notin \Delta$, we let
$\Delta_a$ denote the set $\Delta \cup \Set{a}$ . For a word we $w =
x_0x_1\dots x_{n-1}$ we let $|w| = n$ refer to its length. For $i \in \left[|w|\right]$ we denote by $w[i]$ the symbol on the $i$\textsuperscript{th}
position of $w$ i.e. $w[i] = x_i$. For $a\in\Delta$ and
$w\in\Delta^*$ we let $|w|_a$ denote the number of $a$s in $w$. If $w=v_1v_2$ for some words
$v_1,v_2\in \Delta^*$, then $v_1$ is called a {\em prefix} of $w$ and
$v_2$ is a {\em suffix} of $w$. 
In the remainder of the paper, we let $\Xi = \Sigma \cup \Gamma$ where
$\Sigma$ ($\Gamma$) is a set of symbols called letters (variables) and
$\Sigma\cap \Gamma =\emptyset$. We call a word $w\in\Xi^*$ a \emph{pattern} over $\Xi$. For a pattern $w\in\Xi^*$ we let $\var(w)\subseteq\Gamma$ denote the
set of variables from $\Gamma$ occurring in $w$. A \emph{substitution} for
$\Xi$ is a morphism $S : \Xi^* \to \Sigma^*$ with $S(a) = a$ for every
$a \in \Sigma$ and $S(\epsilon) = \epsilon$. Note, that to define a
substitution $S$, it suffices to define $S(X)$ for all $X\in \Gamma$.   

A \emph{word equation} over $\Xi$ is a tuple $(u,v) \in \Xi^* \times
\Xi^*$ written $u \dotequal v$. A substitution $S$ over $\Xi$ is a
\emph{solution} to a word equation $u\dotequal v$ (denoted $S \subsat u\dotequal v$)   if $S(u) =
S(v)$. A word equation $u \dotequal v$  is
\emph{satisfiable} if there exists a substitution $S$ such that $S
\subsat u\dotequal v$. A \emph{system of word equations} is a set of word equations $P
\subseteq \Xi^* \times \Xi^*$. A system of word equations $P$ is
satisfiable if there exists a substitution $S$ that is a solution to
all word equations (denoted $S \models E$).  \citet{karhumaki2000} showed that for every system of word
equations, a single equation can be constructed which is satisfiable
if and only if the initial formula was satisfiable. The
solution to the constructed word equation can be directly
transferred to a solution of the original word equation system.

\paragraph{Bounded Word Equations}
A natural sub-problem of solving word equations is that of
\emph{Bounded Word Equations}. In this problem we are not only given
a word equation $u \dotequal v$ but also a set of length constraints
$\{\, |X| \leq b_X \mid X \in\Gamma \land b_X \in \mathds{N}\,\}$. The
bounded word equation is satisfiable if there exists a substitution
$S$ such $S\subsat u\dotequal v$ and $|S(X)| \leq b_X$ for each
$X\in\Gamma$. For convenience, we shall sometimes refer to the set of bounds $b_X$ as a function $B: \Gamma \to
\mathds{N}$ such that $b_X = B(X)$.  

\paragraph{Word Equations with Linear Constraints}
A word equation with linear constraints is a word equation $u
\dotequal v$ accompanied by a system $\theta$ of linear Diophantine equations,
where the unknowns correspond to the lengths of possible substitutions
of the variables in $\Gamma$. A word equation with linear constraints is
satisfiable if there exists a substitution $S$ such that $S\subsat
u\dotequal v$ and S satisfies $\theta$. Note that the bounded word
equation problem is in fact a special case of word equations with
linear constraints. 




\paragraph{SAT Solving}
A Boolean formula $\varphi$ with finitely many Boolean variables $\var(\varphi) = \Set{x_1, \dots, x_n}$ is usually given in conjunctive normal form. This is a conjunction over a set of disjunctions (called clauses), i.e. $\varphi = \bigwedge_i \bigvee_j l_{i,j}$, where $l_{i,j} \in \bigcup_{i \in [n]}\Set{x_i, \lnot x_i}$ is a literal. A mapping $\beta : \var(\varphi) \rightarrow \Set{0,1}$ is called an \emph{assignment}; for such an assignment, the literal $l$ evaluates to true if and only if $l = x_i$ and $\beta(x_i) = 1$, or $l = \lnot x_i$ and $\beta(x_i) = 0$. A clause inside a formula in conjunctive normal form is evaluated to true if at least one of its literals evaluates true. We call a formula $\varphi$ \emph{satisfied} (under an assignment) if all clauses are evaluated to true. If there does not exists a satisfying assignment, $\varphi$ is unsatisfiable.

%% file: wordequationsolving.tex
In this section we focus on solving \emph{Bounded Word Equations} and
\emph{Word Equations with Linear Constraints}. We proceed by first
solving bounded word equations, and secondly, we discuss a minor change,
that allows solving word equations with linear constraints. 

\subsection{Solving Bounded Word Equation}
Recall that a bounded word equation consists of a word equation  $u
\dotequal v$ along with a set of equations $\{|X| \leq b_X \}$
providing upper bounds for the solution of each variable $X$s. In our approach we use these bounds to  create  a finite automaton  
which has an accepting run if and only if the bounded word equation is
satisfiable.
 
Before the actual automata construction, we need some convenient
transformations of the word equation itself. For a variable $X$ with length bound
$b_X$,  we replace $X$ with a sequence of new {\em `filled variables'}
$X^{(0)} \cdots X^{(b_X-1)}$ which  we restrict to only be substituted by
either a single letter or the empty word. A pattern containing only
filled variables, as well as letters, is called a {\em filled
  pattern}. For a pattern $w \in \Xi^*$ we denote its corresponding
filled pattern by $\filledWord{w}$.  
In the following, we refer to the alphabet of filled variables by
$\filledWord{\Gamma}$ and by $\filledWord{\Xi} = \Sigma \cup
\filledWord{\Gamma}$ the alphabet of the filled patterns. Let $S :
\Xi^* \rightarrow  \Sigma^*$ be a substitution for $w \in \Xi^*$. We
can canonically define the induced substitution for filled patterns as
$\filledWord{S} : \left(\Sigma \cup \filledWord{\Gamma}\right)
\rightarrow \Sigma_\lambda$ with $\filledWord{S}(a) = S(a)$ for all $a
\in \Sigma$, $\filledWord{S}(X^{(i)}) = S(X)[i]$ for all $X^{(i)} \in
\filledWord{\Gamma}$ and $i < |S(X)|$, and $\filledWord{S}(X^{(j)}) =
\lambda$ for all $X^{(j)} \in \filledWord{\Gamma}$ and $|S(X)| \leq j <
b_X$.  
Here, $\lambda$ is a new symbol ($\lambda \notin \filledWord{\Xi}$) to
indicate an unused position at the end of a filled variable. Note that
the substitution of a single filled variable always maps to exactly
one character from $\Sigma_\lambda$, and, as soon as we discover
$\filledSubs{X^{(j)}} = \lambda$ for $j \in \left[b_X\right]$ it also holds that
$\filledSubs{X^{(i)}} = \lambda$ for all $j\leq i < b_X$. In a sense, the
new element $\lambda$ behaves in the same way as the neutral element
of the word monoid $\Sigma^*$, being actually a place holder for this
element $\epsilon$. In the other direction, if  we have found a satisfying filled
substitution to our word equation, the two filled patterns obtained from the left hand
side and the right hand side of an equation, respectively, we can
transform it to a substitution for our original word equation by defining
$S(X)$ as the concatenation $\filledSubs{X^{(0)}} \dots
\filledSubs{X^{(i)}}$ in which each occurrence of $\lambda$ is
replaced by the empty word $\epsilon$, for all $X \in \Gamma$ and $i
\in [b_X]$.   
	
Our goal is now to build an automaton which calculates a suitable
substitution for a given equation. During the calculation there are
situations where a substitution does not form a total function. To
extend a partial substitution $S : \Xi \pto \Sigma^*$ we define for $X
\in \Xi$ and $b \in \Sigma^*$ the notation $S\subs{X}{b} = S \cup
\Set{X \mapsto b}$ whenever $S(X)$ is undefined and otherwise
$S\subs{X}{b} = S$.	This definition can be naturally applied to filled
substitutions.	  
We define a congruence relation which sets variables and letters in relation whenever their substitution with respect to a partial substitution $\filledWord{S}$ is equal or undefined. 
For all $a,b \in \filledWord{\Xi}\cup\{\lambda\}$ we define 
\fakeMathCenter{$a \congrel b \text{ iff } \filledSubs{a} = \filledSubs{b} \text{ or } \filledSubs{b} \notin \Sigma^*_\lambda \text{ or } \filledSubs{a} \notin \Sigma^*_\lambda.$}	
	\begin{definition}
		\label{def:automata}
		For a word equation $u\dotequal v$ for $u,v \in \Xi^*$ and a mapping $B : \Gamma \rightarrow \mathds{N}$ defining the bounds $B(X)=b_X$, We define the \emph{equation automaton} $A (u\dotequal v,B) = (Q,\delta,I,F)$ where $Q = \left([|\filledWord{u}|+1] \times [|\filledWord{v}|+1]\right) \times \left(\filledLambdaPattern \pto \Sigma_\lambda\right)$ is a set of states consisting of two integers which indicate the position inside the two words $\filledWord{u}$ and $\filledWord{v}$ and a partial substitution, the transition function $\delta : Q \times \Sigma_\lambda \rightarrow Q$ defined by 
			$$\delta\left(\left(\left(i,j\right),  S\right), a\right) = \begin{cases}
				\left(\left(i+1,j+1\right),S\subs{\filledWord{u}[i]}{a}\subs{\filledWord{v}[j]}{a}\right) & \text{if $\filledWord{u}[i] \congrel \filledWord{v}[j] \congrel a$,} \\
				\left(\left(i+1,j\right),S\subs{\filledWord{u}[i]}{\lambda}\right) & \text{if $\filledWord{u}[i] \congrel \lambda = a$,} \\
				\left(\left(i,j+1\right),S\subs{\filledWord{v}[j]}{\lambda}\right) & \text{if $\filledWord{v}[j] \congrel \lambda = a$.}
			\end{cases}$$
		 an initial state $I = \left(\left(0,0\right),\Set{a \mapsto a | a \in \Sigma_\lambda}\right)$ and the set of final states $F = \Set{\left(\left(i,j\right),\filledWord{S}\right) | i = |\filledWord{u}| ,  j = |\filledWord{v}|}$.
		
	\end{definition}
	 The state space of our automaton is finite since the filled substitution $\filledWord{S}$ maps each input to exactly one character in $\Sigma$. The automaton is nondeterministic, as the three choices we have for a transition are not necessarily mutually exclusive.
	
	As an addition to the above definition, we introduce the notion of {\em location} as a pair of integers $(i,j)$ corresponding to two positions inside the two words $\filledWord{u}$ and $\filledWord{v}$. A location $(i,j)$ can also be seen as the set of states of the form $((i,j),S) $ for all possible partial substitutions $S$. 
	
	A run of the above nondeterministic automaton constructs a partial substitution for the given equation which is extended with each change of state. The equation has a solution if one of the accepting states $(|\filledWord{u}|,|\filledWord{v}|,S)$, where $S$ is a total substitution, is reachable, because the automaton simulates a walk through our input equation left to right, with all its positions filled in a coherent way. 
\begin{example}
	Consider the equation $u \dotequal v$ for $u =aZXb , v= aXaY \in  \Xi^*$. We choose the bounds $b_X = b_Y = b_Z = 1$. This will give us the words $\filledWord{u} = aZ^{(0)}X^{(0)}b$ and $\filledWord{v} = aX^{(0)}aY^{(0)}$. \autoref{fig:automaton} visualizes the corresponding automaton. 	
	\begin{figure}[t]
		\centering\resizebox*{!}{7cm}{
		\begin{tikzpicture}[->,>=stealth',shorten >=1pt,auto,node distance=2.5cm,
		semithick]
		\tikzset{every edge/.append style={font=\tiny}}
		\tikzset{every state/.append style={smallState}}
		
		%
		
		\node[pantone280!90] (u1) at (1.7,0.5) {$a$};
		\node[pantone280!90] (u2) at (4.1,0.5) {$Z^{(0)}$};
		\node[pantone280!90] (u3) at (7.1,0.5) {$X^{(0)}$};
		\node[pantone280!90] (u3) at (9.6,0.5) {$b$};
		
		\node[pantone280!90] (v1) at (-0.8,-1.5) {$a$};
		\node[pantone280!90] (v2) at (-0.8,-3.8) {$X^{(0)}$};
		\node[pantone280!90] (v3) at (-0.8,-6.2) {$a$};
		\node[pantone280!90] (v4) at (-0.8,-8.2) {$Y^{(0)}$};
		
		\node[datalabel] (00name)  {$(0,0)$};
		\node[data1]   (00) at($(00name.south)+(0,0.15)$){$S_i$};
		
		\node[node distance=1.4cm] (01name) [right of=00name]  {};
		
		\node[datalabel] (11name) [below of=01name,node distance=1.2cm]  {$(1,1)$};
		\node[data1]   (11) at($(11name.south)+(0,0.15)$){$S_i$};
		
		\node[datalabel] (21name) [right of=11name]  {$(2,1)$};
		\node[data1]   (21) at($(21name.south)+(0,0.15)$){$S_i\subs{Z^{(0)}}{\lambda}$};
		
		\node[datalabel] (12name) [below of=11name,node distance=1.5cm]  {$(1,2)$};
		\node[data1]   (12) at($(12name.south)+(0,0.15)$){$S_i\subs{X^{(0)}}{\lambda}$};
		
		\node[datalabel] (22name) [below of=21name,node distance=1.5cm]  {$(2,2)$};
		\node[data3]   (22) at($(22name.south)+(0,0.15)$){$S_i\subs{Z^{(0)}}{\lambda}\subs{X^{(0)}}{\lambda}$\nodepart{second}$S_i\subs{Z^{(0)}}{a}\subs{X^{(0)}}{a}$\nodepart{third}$S_i\subs{Z^{(0)}}{b}\subs{X^{(0)}}{b}$};
		
		\node[datalabel] (32name) [right of=22name, node distance=3cm]  {$(3,2)$};
		\node[data1]   (32) at($(32name.south)+(0,0.15)$){$S_i\subs{Z^{(0)}}{\lambda}\subs{X^{(0)}}{\lambda}$};
		
		\node[datalabel] (23name) [below of=22name,node distance=2.8cm]  {$(2,3)$};
		\node[data1]   (23) at($(23name.south)+(0,0.15)$){$S_i\subs{Z^{(0)}}{a}\subs{X^{(0)}}{\lambda}$};
		
		\node[datalabel] (33name) [right of=23name, node distance=2.8cm]  {$(3,3)$};
		\node[data2]   (33) at($(33name.south)+(0,0.15)$){$S_i\subs{Z^{(0)}}{a}\subs{X^{(0)}}{a}$\nodepart{second}$S_i\subs{Z^{(0)}}{a}\subs{X^{(0)}}{\lambda}$};
		
		\node[] (34name) [right of=33name]  {};
		
		\node[datalabel] (44name) [below of=34name,node distance=2cm]  {$(4,4)$};
		\node[data2,accepting]   (44) at($(44name.south)+(0,0.15)$){$S_i\subs{Z^{(0)}}{a}\subs{X^{(0)}}{a}\subs{Y^{(0)}}{b}$\nodepart{second}$S_i\subs{Z^{(0)}}{a}\subs{X^{(0)}}{\lambda}\subs{Y^{(0)}}{b}$};

		
		
		\path (00) edge [in=180,out=-90] node {$a$}   (11)
		(11) edge [in=90,out=-90] node {$\lambda$}   (12name)
		edge [in=180,out=0] node {$\lambda$}   (21)
		edge [in=150,out=-10] node[xshift=0cm,yshift=-0.5cm] {$\lambda$}   (22)
		edge [in=180,out=-22] node[xshift=-0.05cm,yshift=-0.4cm] {$a$}  (22)
		edge [in=210,out=-30] node[xshift=0cm,yshift=-0.8cm] {$b$}  (22)
		(12) edge [in=180,out=-90] node {$a$}   (23)    
		
		(21)edge [in=180,out=0] node {$\lambda$}   (32)
		(22) edge [in=160,out=0] node {$a$}   (33)
		(23)edge [in=200,out=0] node[xshift=-0.32cm,yshift=-0.2cm] {$\lambda$}   (33)
		(33)edge [in=140,out=10] node {$b$}   (44)
		edge [in=190,out=-80] node {$b$}   (44)
		;
		
		\end{tikzpicture}}
		\caption{\label{fig:automaton}Automaton for the word equation $aZXb \dotequal aXaY$, with the states grouped according to their locations. Only reachable states are shown.}
		\vspace*{-0.5cm}
	\end{figure}
	A run starting with the initial substitution $S_i = \Set{a \mapsto a | a \in \Sigma_\lambda}$ reaching one of the final states gives us a solution to the equation. In this example we get the substitutions $Z \mapsto a, X \mapsto a, Y \mapsto b$ and  $Z \mapsto a, X \mapsto \epsilon, Y \mapsto b$.
\end{example}
	\begin{theorem}
		Given a bounded word equation $u \dotequal v$ for $u,v \in
        \Xi^*$,with bounds $B$, then the
        automaton $A (u \dotequal v,B)$  reaches an accepting state  if and only if there exists $S$ such that $S
        \subsat u \dotequal v$ and $|S(X)| \leq B(X)$ for all $X \in \Gamma$.
	\end{theorem}
	
    \subsubsection*{SAT Encoding}
    We now encode the solving process into propositional logic. For that we impose an ordering on the finite alphabets $\Sigma = \Set{a_0, \dots, a_{n-1}}$ and $\Gamma = \Set{X_0, \dots, X_{m-1}}$ for $n, m \in \mathds{N}$. Using the upper bounds given for all variables $X \in \Gamma$, we create the filled variables alphabet $\filledWord{\Gamma}$. Further, we create the Boolean variables $\variableVars^a_{X^{(i)}}$, for all $X^{(i)} \in \filledWord{\Gamma}$, $a \in \Sigma_\lambda$ and $i \in [b_X]$. Intuitively, we want to construct our formula such that an assignment $\beta$ sets $\variableVars^a_{X^{(i)}}$ to $1$, if the solution of the word equation, which corresponds to the assignment $\beta$, is such that at position $i$ of the variable $X$ an $a$ is found, meaning $\filledSubs{X^{(i)}} = a$.
	  To make sure $\variableVars^a_{X^{(i)}}$  is set to $1$ for exactly one $a \in \Sigma_\lambda$ we define the clause $\bigvee_{a \in \Sigma_\lambda} \variableVars^a_{X^{(i)}}$ which needs to be assigned true, as well the constraints $\variableVars^a_{X^{(i)}} \rightarrow \lnot \variableVars^b_{X^{(i)}}$, for all $a,b \in \Sigma_\lambda, X \in \Gamma, i \in [b_X]$ where $a \neq b$, which also need to be all true.
	 
	 To match letters we add the variables $\constsVars_{a,a} \leftrightarrow \top$ and $\constsVars_{a,b} \leftrightarrow \bot$ for all $a,b \in \Sigma_\lambda$ with $a \neq b$. As such, the actual encoding of our equation can be defined as follows: for $w \in \Set{\filledWord{u},\filledWord{v}}$ and each position $i$ of $w$ and letter $a \in \Sigma_\lambda$ we introduce a variable which is true if and only if $w[i]$ will correspond to an $a$ in the solution of the word equation. More precisely, we make a distinction between constant letters and variable positions and define: $\wordEncoding_{w[i]}^a \leftrightarrow \constsVars_{w[i],a}$ if $w[i] \in \Sigma_\lambda$ and $\wordEncoding_{w[i]}^a \leftrightarrow \variableVars^a_{w[i]}$ if $w[i] \in \filledWord{\Gamma}$.	 
	 The equality of two characters, corresponding to position $i$ in $u$ and, respectively, $j$ in $v$, is encoded by introducing a Boolean variable $\wordsMatch_{i,j} \leftrightarrow  \bigvee_{a \in \Sigma_\lambda} \wordEncoding_{u[i]}^a \land \wordEncoding_{v[j]}^a$ for appropriate $i \in [|\filledWord{u}|]$, $j \in [|\filledWord{v}|]$.
	 
	 Based on this setup, each location of the automaton is assigned a
     Boolean variable. As seen in \autoref{def:automata} we process
     both sides of the equation simultaneously, from left to
     right. As such, for a given equation $u \dotequal v$ we create $n\cdot m=\left(|\filledWord{u}|+1\right) \cdot \left(|\filledWord{v}|+1\right)$ many Boolean variables $\stateVars_{i,j}$ for $i \in [n]$ and $j \in [m]$. Each variable corresponds to a location in our automaton. The location $(0,0)$ is our initial location and $(|\filledWord{u}|,|\filledWord{v}|)$ our accepting location. The goal is to find a path between those two locations, or, alternatively, a satisfying assignment $\beta$, which sets the variables corresponding to these locations to $1$. Every path between the location $(0,0)$ and another location corresponds to matching prefixes of $u$ and $v$, under proper substitutions. We will call locations where an assignment $\beta$ sets a variable $\stateVars_{i,j}$ to $1$, active locations. Our transitions are now defined by a set of constraints. We fix $i \in [n]$ and $j \in [m]$ in the following. 
	 The constraints are given as follows:
	 	  The first constraint (\ref{eq:1}) ensures that every active location has at least one active successor. The next three constraints (\ref{eq:2})-(\ref{eq:4}) ensure the validity of the paths we follow: from a location we can  only proceed to exactly one other location, in order to find a satisfying assignment; therefore we disallow simultaneous steps in multiple directions. In (\ref{eq:5}),(\ref{eq:6}) we forbid using an $\lambda$-transition whenever there is another possibility of moving forward. In the same manner we proceed in the case of two matching $\lambda$ in (\ref{eq:7}); this part is especially important for finding substitutions which are smaller than the given bounds. The idea applies in the same way for matching letters, whose encoding is given in (\ref{eq:8}). The actual transitions which are possible from one state to another are encoded in (\ref{eq:9}) by using our Boolean variables $\wordsMatch_{i,j}$ which are true for matching positions in the two sides of the equation. This constraint allows us to move forward in both words if there was a match of two letters in the previous location. When the transitions are pictured as movements in the plane, this corresponds to a diagonal move. A horizontal or vertical move corresponds to a match with the empty word. The last constraint (\ref{eq:10}) ensures a valid predecessor. This is supposed to help the solver in deciding the satisfiability of the obtained formula, i.e., to guide the search in an efficient way. It can be seen as a local optimization step.	 
	  	 \begin{align}
	  &\stateVars_{i,j} \rightarrow \stateVars_{i+1,j} \lor \stateVars_{i,j+1} \lor \stateVars_{i+1,j+1}\label{eq:1}\\
	  &\left(\stateVars_{i,j} \land \stateVars_{i,j+1}\right) \rightarrow \left(\lnot \stateVars_{i+1,j+1} \land \lnot \stateVars_{i+1,j}\right)\label{eq:2}\\
	  &\left(\stateVars_{i,j} \land \stateVars_{i+1,j}\right) \rightarrow \left(\lnot \stateVars_{i+1,j+1} \land  \lnot \stateVars_{i,j+1}\right)\label{eq:3}\\
	  &\left(\stateVars_{i,j} \land \stateVars_{i+1,j+1}\right) \rightarrow\left( \lnot \stateVars_{i,j+1} \land \lnot \stateVars_{i+1,j}\right)\label{eq:4}\\
	  &\stateVars_{i,j} \land \lnot\wordEncoding_{u[i]}^\lambda \rightarrow \lnot \stateVars_{i+1,j}\text{ and }\stateVars_{i,j} \land \wordEncoding_{u[i]}^\lambda \land \lnot\wordEncoding_{v[j]}^\lambda \rightarrow \stateVars_{i+1,j}\label{eq:5}\\
	  &\stateVars_{i,j} \land \lnot\wordEncoding_{v[j]}^\lambda \rightarrow \lnot \stateVars_{i,j+1}\text{ and }\stateVars_{i,j} \land \lnot\wordEncoding_{u[i]}^\lambda \land \wordEncoding_{v[j]}^\lambda \rightarrow \stateVars_{i,j+1}\label{eq:6}\\
	  &\stateVars_{i,j} \land \wordEncoding_{u[i]}^\lambda \land \wordEncoding_{v[j]}^\lambda \rightarrow \stateVars_{i+1,j+1}\label{eq:7}\\
	  &\stateVars_{i,j} \land \stateVars_{i+1,j+1} \rightarrow \wordsMatch_{i,j} \label{eq:8}\\
	  &\stateVars_{i,j} \leftrightarrow \left(\stateVars_{i-1,j-1} \land \wordsMatch_{i-1,j-1} \right) \lor \left(\stateVars_{i,j-1} \land \lnot \wordsMatch_{i,j-1} \right)  \lor \left(\stateVars_{i-1,j} \land \lnot \wordsMatch_{i-1,j} \right)\label{eq:9}\\
	  &\stateVars_{i+1,j+1} \rightarrow \stateVars_{i,j} \lor \stateVars_{i+1,j} \lor \stateVars_{i,j+1}\label{eq:10}
	  \end{align}
	 The final formula is the conjunction of all constraints defined above. This formula is true iff location $(n,m)$ is reachable from location $(0,0)$, and this \begin{wrapfigure}[12]{l}{0.33\textwidth} 
	 	\vspace*{-0.5cm}
	 	\resizebox*{!}{3.4cm}{
	 		\begin{tikzpicture}[rotate=-90]
	 		\foreach \x in {0,...,37}
	 		\foreach \y in {0,...,31}
	 		{
	 			\draw[pantone280!30] (\y,\x) +(.5,.5) rectangle ++(-.5,-.5);
	 		}
	 		
	 		\foreach \point in
	 		{(0,0),(0,1),(1,0),(1,1),(1,2),(2,1),(2,2),(2,3),(2,4),(2,5),(2,6),(2,7),(2,8),(3,1),(3,2),(3,3),(3,4),(3,5),(3,6),(3,7),(3,8),(3,9),(3,10),(4,1),(4,2),(4,3),(4,4),(4,5),(4,6),(4,7),(4,8),(4,9),(4,10),(4,11),(4,12),(5,1),(5,2),(5,3),(5,4),(5,5),(5,6),(5,7),(5,8),(5,9),(5,10),(5,11),(5,12),(5,13),(5,14),(6,1),(6,2),(6,3),(6,4),(6,5),(6,6),(6,7),(6,8),(6,9),(6,10),(6,11),(6,12),(6,13),(6,14),(6,15),(6,16),(7,1),(7,2),(7,3),(7,4),(7,5),(7,6),(7,7),(7,8),(7,9),(7,10),(7,11),(7,12),(7,13),(7,14),(7,15),(7,16),(7,17),(7,18),(8,1),(8,2),(8,3),(8,4),(8,5),(8,6),(8,7),(8,8),(8,9),(8,10),(8,11),(8,12),(8,13),(8,14),(8,15),(8,16),(8,17),(8,18),(8,19),(8,20),(9,2),(9,3),(9,4),(9,5),(9,6),(9,7),(9,8),(9,9),(9,10),(9,11),(9,12),(9,13),(9,14),(9,15),(9,16),(9,17),(9,18),(9,19),(9,20),(9,21),(10,3),(10,4),(10,5),(10,6),(10,7),(10,8),(10,9),(10,10),(10,11),(10,12),(10,13),(10,14),(10,15),(10,16),(10,17),(10,18),(10,19),(10,20),(10,21),(11,10),(11,11),(11,12),(11,13),(11,14),(11,15),(11,16),(11,17),(11,18),(11,19),(11,20),(11,21),(11,22),(12,11),(12,12),(12,13),(12,14),(12,15),(12,16),(12,17),(12,18),(12,19),(12,20),(12,21),(13,12),(13,13),(13,14),(13,15),(13,16),(13,17),(13,18),(13,19),(13,20),(13,21),(13,22),(14,13),(14,14),(14,15),(14,16),(14,17),(14,18),(14,19),(14,20),(14,21),(15,16),(15,17),(15,18),(15,19),(15,20),(15,21),(15,22),(16,17),(16,18),(16,19),(16,20),(16,21),(17,18),(17,19),(17,20),(17,21),(17,22),(18,19),(18,20),(18,21),(18,22),(18,23),(19,22),(19,23),(19,24),(19,25),(19,26),(19,27),(19,28),(20,23),(20,24),(20,25),(20,26),(20,27),(20,28),(20,29),(21,24),(21,25),(21,26),(21,27),(21,28),(21,29),(21,30),(22,29),(22,30),(22,31),(22,32),(22,35),(23,30),(23,31),(23,32),(23,33),(23,35),(24,30),(24,31),(24,32),(24,33),(24,34),(24,35),(25,30),(25,31),(25,32),(25,33),(25,34),(25,35),(25,36),(26,35),(26,36),(26,37),(27,36),(27,37),(28,36),(29,36),(30,36),(31,37)}
	 		{
	 			\node[consideredState,rectangle,minimum width=1cm, minimum height=1cm] () at \point { };
	 		}
	 		
	 		\foreach \point in
	 		{(0,0),(1,1),(2,2),(3,3),(4,4),(5,5),(6,6),(7,7),(8,8),(9,9),(10,10),(11,11),(12,12),(13,13),(13,14),(13,15),(14,16),(15,17),(16,18),(17,19),(17,20),(17,21),(18,22),(19,23),(20,24),(20,25),(20,26),(20,27),(20,28),(21,29),(22,30),(23,31),(24,32),(24,33),(24,34),(25,35),(26,36),(27,37),(28,37),(29,37),(30,37),(31,37)}
	 		{
	 			\node[nonconsideredstate,rectangle,minimum width=1cm, minimum height=1cm] () at \point { };
	 		}
	 		\end{tikzpicture} }
	 	\captionof{figure}{Solver~computation on $XaXbYbZ \dotequal aXY\-YbZZbaa$}
	 	\label{fig:automatoncomp}
	 \end{wrapfigure}is true iff the given word equation is satisfiable w.r.t. the given length bounds.
	 
	 \begin{lemma}
	 	Let $u \dotequal v$ be a word equation, $B$ be the function giving the bounds for the word equation variable, and $\varphi$ the corresponding formula consisting of the conjunction (\ref{eq:1}) - (\ref{eq:10}) and the earlier defined constraints in this section, then  $\varphi \land \stateVars_{0,0} \land \stateVars_{|\filledWord{u}|,|\filledWord{v}|}$ has a satisfying assignment if and only if $A(u \dotequal v,B)$ reaches an accepting state.
	 \end{lemma}

 	\begin{example}
	 Consider the word equation $u \dotequal v$ where $u = XaXbYbZ$ and $v = aXYYbZZbaa \in \Xi^*$ where $\Sigma = \Set{a}$ and $\Gamma = \Set{X,Y,Z}$. Using the approach discussed above, we find the solution $S(X) = aaaaaaaa$, $S(Y) = aaaa$ and $S(Z) = aa$ using the bounds $b_X = 8$ and $b_Y = b_Z = 6$. We set up an automaton with $32 \cdot 38 = 1216$ states to solve the equation. In \autoref{fig:automatoncomp} we show the computation of the SAT-Solver. Light grey markers indicate states considered in a run of the automaton. In this case only $261$ states are needed. The dark grey markers visualize the actual path in the automaton leading to the substitution. Non-diagonal stretches are $\lambda$ transitions. 
	 \end{example} 
	
	\subsection{Refining Bounds and Guiding the Search} \label{sub:rb}
    Initial experiments revealed a
    major inefficiency of our approach: most of the locations  were not used during the search and only increased the encoding
    time. The many white markers in \autoref{fig:automatoncomp} indicating unused locations visualizes the problem.
    Since we create all required variables $x \in \Gamma$ and constraints for every
    position $i < b_X$,   we can reduce the automaton size by lowering
    these upper bounds.  Abstracting a word equation by the length of the variables gives us a way to refine the bounds $b_X$ for some of the variables $X \in \Gamma$. By only considering length we obtain a Diophantine equation in the following manner. We assume an ordering on the variable alphabet $\Gamma = \Set{X_0, \dots, X_{n-1}}$. We associate to each word equation variable $X_j$ an integer variable $I_j$. 
	\begin{definition}\label{lem:abs-len}
      For a word equation $u \dotequal v$ with $\Gamma = \Set{X_0,\dots,X_{n-1}}$  we define its \emph{length abstraction} by
      $\sum_{j \in [n]} \left(|u|_{X_j} - |v|_{X_j}\right) \cdot I_j = \sum_{a \in \Sigma} |v|_a - |u|_a $ 	 for $j \in [n]$.	
	\end{definition}If a word equation has a solution $S$, then so does
    its length abstraction with variable $I_j =
    |S(X_j)|$.  Our interest is computing upper bounds
    for each variable $X_k \in \Gamma$ relative to the upper
    bounds of the bounded word equation problem. To this end consider
    the following natural deductions
    \begin{align*}
      &\sum_{j \in [n]} \left(|u|_{X_j} - |v|_{X_j}\right) \cdot I_j
     	= \sum_{a \in \Sigma} \left(|v|_a - |u|_a\right)\\
      \Longleftrightarrow~& I_k = \frac{\sum_{a \in \Sigma}\left( |v|_a - |u|_a\right)}{|u|_{X_k}-|v|_{X_k}} - \frac{\sum_{j\in
              [n]\setminus k}  \left(|u|_{X_j} - |v|_{X_j}\right)\cdot
              I_j}{|u|_{X_k}-|v|_{X_k}}\\
      \Longrightarrow~& I_k \leq \frac{\sum_{a \in \Sigma} |v|_a -
      |u|_a}{\left(|u|_{X_k}-|v|_{X_k}\right)} - \frac{\sum_{j\in \kappa}  \left(|u|_{X_j} - |v|_{X_j}\right)\cdot
              b_{X_j}}{\left(|u|_{X_k}-|v|_{X_k}\right)} = b^\mathsf{S}_{X_k},                 
    \end{align*}
    where $\kappa = \set{m \in[n]\setminus{k} \mid (|u|_{X_k}-|v|_{X_k}) \cdot (|u|_{X_m}-|v|_{X_m}) < 0 }$. Whenever
    $0 < b^\mathsf{S}_{X_k} < b_{X_k}$ holds, we use $b^\mathsf{S}_{X_k}$ instead
    of $b_{x_k}$ to prune the search space. 

    The length abstraction is also useful because it might 
    give information about the unsatisfiability of an
    equation: if there is no
    solution to the Diophantine equation, there is no solution to the
    word equation. We use this acquired knowledge and directly
    report this fact. 
	Unfortunately whenever $|u|_{X}-|v|_{X} = 0$ holds for a variable $X$ we cannot refine the bounds, as they are not influenced by the above Diophantine equation.
	
	\paragraph{Guiding the Search}
    The length abstraction used to refine upper bounds can also be
    used to guide the search in the automaton. In particular it
    can restrict allowed length of one variable based on the
    length of others. We refer to the coefficient of variable
    $I_j$  in \autoref{lem:abs-len}  by $\coefficients(X_j) = \left(|u|_{X_j} -
      |v|_{X_j}\right)$. 

    To benefit from the abstraction of the
    word equation inside our propositional logic encoding we use
    Reduced Ordered Multi-Decision Diagrams (MDD)
    \cite{abio2014encoding}. An MDD is a directed acyclic graph, with two nodes having no outgoing edges (called
    \texttt{true} and \texttt{false} terminal nodes). A Node in the MDD  is associated to exactly one variable
    $I_j$, and has an
    outgoing edge for each element of $I_j$s
    domain. In the MDD, a node labelled $I_j$ is only connected to nodes labelled
    $I_{j+1}$. A row ($\mathsf{r}(I_j)$) in an MDD is a  a subset of nodes
    corresponding to a certain variable $I_j$. 	
	
	We create the MDD following 
    \autoref{lem:abs-len}. The following definition creates the
    rows of the MDD recursively. An MDD node is
    a tuple consisting of a variable $I_j$ and an integer
    corresponding the partial sum which can be obtained using
    the coefficients and position information of all previous
    variables $I_k$ for $k < j$.  We introduce a new
    variable $I_{-1}$ labelling the initial node of
    the MDD. The computation is done as follows:   
	\begin{align}
	\mathsf{r}(I_i) = \{\;(I_i,s+k\cdot \coefficients(X_i))\;|\;& s \in \Set{s' | (I_{i-1},s') \in \mathsf{r}(I_{i-1})} , k \in [b_{X_i}]\;\}
	\end{align}
	and $\mathsf{r}(I_{-1}) = \Set{(I_{-1},0)}$. Since $I_j$ is associated to  word equation $X_j$, we let $\mathsf{r}(X_{j}) = \mathsf{r}(I_{j})$.  The whole set of nodes in the MDD we denote by $M^C = \bigcup_{X
      \in \Gamma \cup\Set{X_{-1}}} \mathsf{r}(X)$. The $\mathtt{true}$ node of the MDD is $(I_{n-1}, \mathsf{s_\#})$,
    where $ \mathsf{s_\#} = \sum_{a \in \Sigma} |v|_a -
    |u|_a$. If the initial creation of nodes did not
    add this node, the given equation (\autoref{lem:abs-len}) is not satisfiable hence
    the word equation has no solution given the set bounds.  
    Furthermore there is no need to encode the full MDD, when only a
    subset of its nodes can reach $(I_{n-1},\mathsf{s_\#} )$.
    For reducing the MDD nodes to this subset, we calculate all
    predecessors of a given node $(I_i,s) \in M^C$ as
    follows \[\mathsf{pred}((I_i,s)) = \Set{(I_{i-1},s') | s' = s - k
        \cdot \coefficients(X_{i-1}) , k \in [b_{X_{i-1}}]}.\] The
    minimized set $M = F(T)$ of reachable nodes starting at the only
    accepting node $T = \Set{(I_{n},\mathsf{s_\#})}$ is afterwards
    defined through a fixed point by   
	\begin{align}
		T \subseteq F(T) \land\left(\forall\; p \in F(T) : q \in \mathsf{pred}(p) \land q \in M^C \Rightarrow q \in F(T)\right)\label{eq:fixpoint}
	\end{align}	
	We continue by encoding this into a Boolean formula. For
    that we need information on the actual length of a possible
    substitution. We reuse the Boolean variables of our
    filled variables $X \in \filledWord{\Gamma}$. The idea is to
    introduce $b_X+1$ many Boolean variables $(\onehot_i(0) \dots
    \onehot_i(b_X+1)) $ for each $X_i \in
    \Gamma$, where $\onehot_i(j)$ is true if and only if $X_i$ has
    length $j$ in the acutal substitution. 
    To achieve this we add a constraint forcing substitutions
    to have all $\lambda$ in the end. 
    We force
    our solver to adapt to this by adding clauses
    $\variableVars^{\lambda}_{X^{(j)}} \rightarrow
    \variableVars^{\lambda}_{X^{(j+1)}}$ for all $j \in [b_{X_i}-1]$ and
    $X_i^{(j)} \in \filledWord{\Gamma}$. The actual encoding is done by
    adding the constraints  
	$\onehot_i(0) \leftrightarrow
    \variableVars^{\lambda}{X_i^{(0)}}$ and $\onehot_i(b_{X_i})
    \leftrightarrow
    \lnot\variableVars^{\lambda}_{X^{(b_{X_i}-1)}}$,   
	which fixes the edge cases for the substitution by the empty word and no $\lambda$ inside it. For all $j \in [b_{X_i}]_0$, we add the constraints $\onehot_i(j)\leftrightarrow \variableVars^{\lambda}_{X_i^{(j)}} \land \lnot \variableVars^{\lambda}_{X_i^{(j-1)}}$, which marks the first occurrence of $\lambda$.
%

	The encoding of the MDD is done nodewise by associating a Boolean variable $\mddVar_{i,j}$ for each $i \in [|\Gamma|]$, where $(I_i,j) \in M$. Our goal is now to find a path inside the MDD from node $(I_{-1},0)$ to $(I_{n-1},\mathsf{s_\#})$. Therefore we  enforce a true assignment for the corresponding variables $\mddVar_{-1,0}$ and $\mddVar_{{n-1},\mathsf{s_\#}}$. 
	A valid path is encoded by the constraint $\mddVar_{i-1,j} \land \onehot_i(k) \rightarrow \mddVar_{i,s}$ for each variable $X_i \in \Gamma$, $k \in [b_{X_i}]_0$, where $s = j + k \cdot \coefficients(X_i)$ and $(I_i,s) \in M$. This encodes the fact that whenever we are at a node $(I_{i-1},s) \in M$ and the substitution for a variable $X_i$ has length $k$ ($|S(X_i)| = k$), we move on to the next node, which corresponds to $X_{i}$ and an integer obtained by taking the coefficient of the variable $X_i$, multiplying it by the substitution length,
	\begin{wrapfigure}[13]{r}{0.38\textwidth} 
		\vspace*{-0.6cm}
		\centering
		\resizebox*{3.9cm}{!}{
			\begin{tikzpicture}[->,>=stealth',shorten >=1pt,auto,node distance=1.9cm,
			semithick]
			\tikzset{every edge/.append style={font=\tiny}}
			\tikzset{every state/.append style={bigState,rectangle,rounded corners=3,minimum size=0.6cm,font=\tiny}}
			\node[initial,state]   (0)              {$(X_{-1},0)$};
			\node[state] (1) [below of=0, node distance=1.7cm] {$(X_1,0)$};
			\node[state] (3) [below of=1, node distance=1.35cm] {$(X_2,-1)$};
			\node[state] (2) [left of=3,node distance=1cm,anchor=east,draw=pantone280,fill=pantone280!15,text=pantone280] {$(X_2,0)$};
			\node[state] (4) [right of=3,node distance=1cm,anchor=west,draw=pantone280,fill=pantone280!15,text=pantone280] {$(X_2,-2)$};
			\node[state] (5) [below of=3, node distance=1.35cm] {$(X_3,0)$};

			\path (0) edge [] node [fill=white, anchor=center] {\parbox{0.9cm}{$0 + 0 \cdot 0$\\\color{pantone280}$0 + 1 \cdot 0$\\$0 + 2 \cdot 0$}}   (1)
			(1) edge [bend right,lightText,anchor=west] node [anchor=east] {$0 + 0 \cdot -1$}   (2)
			edge [] node [fill=white, anchor=center] {$0 + 1 \cdot -1$}   (3)
			edge [bend left,lightText] node [near start,right]  {$0 + 2 \cdot -1$}   (4)
			(2) edge [bend right,lightText] node [xshift=-0cm,yshift=0cm,anchor=east] {$0 + 0 \cdot 1$}   (5)
			(3) edge [] node [fill=white, anchor=center] {$-1 + 1 \cdot 1$}   (5)
			(4) edge [bend left,lightText] node [anchor=west] {$-2 + 2 \cdot 1$}   (5)
			;
			\end{tikzpicture}}
		\captionof{figure}{The MDD for  $aX_1aX_2 \dotequal aX_3X_1b$}
		\label{fig:mdd}
	\end{wrapfigure}%
 and adding it to the previous partial sum $s$. Whenever there is only one successor to a node $(I_i,j)$ within our MDD, we directly force its corresponding one hot encoding	
 to be true by adding $\mddVar_{i-1,j} \rightarrow \onehot_i(j)$. This reduces the amount of guesses on variables.
		\begin{example}
			Consider the equation $u \dotequal v$ for $u = aX_1aX_2, v= aX_3X_1b \in  \Xi^*$, where $\Sigma = \Set{a,b}$ and $\Gamma = \Set{X_1,X_2,X_3}$. The corresponding linear equation therefore has the form $0\cdot I_1-I_2+I_3 = 0$ which gives us the coefficients $\coefficients(X_1) = 0$, $\coefficients(X_2) = -1$ and $\coefficients(X_3) = 1$. For given bounds $b_{X_1} = b_{X_2} = b_{X_3} = 2$ the reduced MDD has the form shown in \autoref{fig:mdd}. We only draw transitions which are transitively connected to the \texttt{true} terminal node. The solver returns the substitution $S(X_1) = \epsilon$, $S(X_2) = b$ and $S(X_3) = a$. Therefore it took the path marked in dark grey inside the MDD.
		\end{example}

        \paragraph{Adding Linear Length Constraints} \label{sub:lc}
        Until now we have only concerned ourselves with bounded
        word equations. As mentioned in the introduction however,
        bound\-ed equations with linear constraints are of interest as
        well. In particular, without loss of generality we restrict to linear constraints
        of the form~\citep{abio2014encoding} $c_0 I_0 + \cdots + c_{n-1} I_{n-1} \leq c$ where
        $c, c_i \in \mathds{Z}$ are integer coefficents and $I_i$ are
        integer variables with a domain $D_i = \Set{m \in \mathds{Z} |
          -d_i \leq m \leq d_i}$ and a corresponding
        $d_i\in\mathds{Z}$. 
        Each $I_i$
        corresponds to the length of a substitution to a variable of the given word equation.
        
        Notice that the structure of the linear length constraint is 
        similar to that of \autoref{lem:abs-len}. For handling
        linear constraints we can  adapt the generation of MDD
        nodes to keep track the partial sum of the linear constraint, 
        and define the accepting node of the MDD to one where all rows
        have been visited and the inequality is true. We simply extend the set
        $T$ which was used in the fix point iteration in
        (\ref{eq:fixpoint}) to the set $T = \Set{(I_n,s) |
        (I_n,s) \in M^C \land s \leq \mathsf{s_\#}}$. 



%% file: experiments.tex
	
\begin{figure}[t]
		\centering
			\resizebox{0.75\textwidth}{!}{
				\begin{tikzpicture}[->,>=stealth',shorten >=1pt,auto,node distance=3cm,
				semithick]
				\tikzset{every edge/.append style={font=\footnotesize,draw=black}}
				\tikzset{every state/.append style={rectangle,minimum width=2cm,anchor=west,rounded corners=3pt, text width=2cm,align=center,font=\footnotesize}}
				
				\draw[black,rounded corners=3pt,fill=black!0] (-0.5,-1)  rectangle (8.75,1);
				\draw[black,rounded corners=3pt,fill=black!3] (2.5,-0.75)  rectangle (8.5,0.75);
				
				\node[state,draw=black,color=black] (0) at (0,0) {{\large \faCogs}\\Preprocessing};
				\node[state,draw=black,color=black,fill=black!5] (1) [right of=0]   {{\large \faWrench}\\Encoding};
				\node[state,draw=black,color=black,fill=black!5] (2) [right of=1]  {{\large \faPuzzlePiece}\\SAT-Solver};
				\node[] (3) [left of=0] {Input};
				\node[] (4) [] at (10,-.5)  {Model};
				\node[] (w) [black] at (0.8,1.1) {\small\woorpje};

				\path (3) edge [] node {} (0)
				(0) edge [anchor=north east] node {\texttt{unsat}}   (.5,-1.75)
				edge [out=-90,in=-120,looseness=0.5,font=\footnotesize] node {\texttt{sat}}   (4)
				edge [font=\footnotesize] node {} (1)
				(1) edge [] node {} (2)
				(2) edge [] node {\texttt{sat}}   (4)
				(5.5,0.75) edge [out=120,in=60,looseness=30]  node {\texttt{unsat}} (5.6,0.75)
				(8.5,0.5) edge [anchor=south west]  node {\texttt{unknown}} (10,0.5);
				\end{tikzpicture}
			}
		\caption{Architecture of \woorpje}		\label{fig:architecture}
		\vspace*{-0.3cm}
	\end{figure}
    The approach described  in the previous sections has been
    implemented in the tool \woorpje. The inner workings of \woorpje\
    is visualised in \autoref{fig:architecture}. \woorpje\ first has a
    preprocessing step where obviously satisfiable/unsatisfiable word equations
    are immediately reported. 
    After the preprocessing step, \woorpje\ iteratively encodes the
    word equation into a propositional logic formula and solves it with Glucose~\citep{audemard2018glucose} for increasing maximal
    variable lengths ($i^2$, where $i$ is the current iteration). If a solution is found, it is reported. The maximal value of $i$ is user
    definable, and by default set to $2^n$ where $n$ is the length of the given equation. If \woorpje\ reaches the given bound without a verdict, it returns
    unknown.


	We have run \woorpje{} and state of the art word
    equation solvers (CVC4 1.6, Norn 1.0.1, Sloth 1.0, Z3 4.8.4) on several
    word equation benchmarks with linear length constraints. The benchmarks range from theoretically-interesting cases to
    variations of the real-world application set Kaluza \cite{saxena2010symbolic}. 
All tests were performed on Ubuntu Linux 18.04 with an Intel Xeon E5-2698 v4 @ 2.20GHz CPU and 512GB of memory with a timeout of 30 seconds.	

	We used five different kind of benchmarks. The first track (I) 
	was produced by generating random strings, and replacing factors with variables at random, in a coherent fashion. This guarantees the existence of a solution.
	The generated word equations have at most 15 variables, 10 letters, and length 300. The second track (II) is based on the idea in Proposition 1 of \cite{MFCS2017},  where the equation  
	$X_naX_nbX_{n-1}bX_{n-2}\cdots bX_1\doteq aX_nX_{n-1}X_{n-1}bX_{n-2}X_{n-2}b \cdots b X_{1}X_{1}baa$
	is shown to have a minimal solution of exponential length w.r.t. the length of the equation. The third track (III) is based on the second track, but each letter $b$ is
	 replaced by the left hand side or the right hand side of some randomly generated word equation (e.g., with the methods from track (I)). In the fourth track (IV) each benchmark consists of a system of 100 small random word equations with at most 6 letters, 10 variables and length 60. The hard aspect of this track is solving multiple equations at the same time. Within the fifth track (V) each benchmark enriches a system of 30 word equations by suitable linear constraints, as presented in this paper. This track is inspired by the Kaluza benchmark set 
	 in terms of having many small equations enriched by linear length constraints. All tracks, except track II which holds 9 instances, consist of 200 benchmarks. The full benchmark set is available at \url{https://www.informatik.uni-kiel.de/~mku/woorpje}. 
		\begin{table}[t]
		\centering
		\resizebox{\textwidth}{!}{
			\begin{tabular}{*{30}{c}}
				& \multicolumn{5}{c}{\textsc{Track I}} & & \multicolumn{5}{c}{\textsc{Track II}}& & \multicolumn{5}{c}{\textsc{Track III}}& & \multicolumn{5}{c}{\textsc{Track IV}} & &\multicolumn{5}{c}{\textsc{Track V}} \\
				\cmidrule(lr){2-6}\cmidrule(lr){8-12}\cmidrule(lr){14-18}\cmidrule(lr){20-24}\cmidrule(lr){26-30}
				& \tiny \faCheckCircleO & \tiny \faStopCircleO & \tiny \faPauseCircleO & \tiny \faTimesCircleO  & \tiny\faClockO & & \tiny \faCheckCircleO & \tiny \faStopCircleO & \tiny \faPauseCircleO & \tiny \faTimesCircleO  & \tiny\faClockO & & \tiny \faCheckCircleO & \tiny \faStopCircleO & \tiny \faPauseCircleO & \tiny \faTimesCircleO  & \tiny\faClockO & & \tiny \faCheckCircleO & \tiny \faStopCircleO & \tiny \faPauseCircleO & \tiny \faTimesCircleO  & \tiny\faClockO & & \tiny \faCheckCircleO & \tiny \faStopCircleO & \tiny \faPauseCircleO & \tiny \faTimesCircleO  & \tiny\faClockO\\
				\cmidrule(lr){2-6}\cmidrule(lr){8-12}\cmidrule(lr){14-18}\cmidrule(lr){20-24}\cmidrule(lr){26-30}
				\tiny\scshape woorpje& \tiny200 & \tiny0 & \tiny0 & \tiny0 & \tiny8.10&& \tiny5 & \tiny0 & \tiny4 & \tiny0 & \tiny123.85&& \tiny189 & \tiny0 & \tiny11 & \tiny0 & \tiny341.74&& \tiny196 & \tiny2 & \tiny2 & \tiny0 & \tiny136.20&& \tiny178 & \tiny9 & \tiny13 & \tiny0 & \tiny399.22\\
				\tiny\scshape cvc4& \tiny182 & \tiny0 & \tiny17 & \tiny0 & \tiny543.32&& \tiny1 & \tiny0 & \tiny8 & \tiny0 & \tiny240.03&& \tiny165 & \tiny0 & \tiny35 & \tiny0 & \tiny1055.24&& \tiny172 & \tiny0 & \tiny28 & \tiny0 & \tiny925.13&& \tiny179 & \tiny0 & \tiny21 & \tiny0 & \tiny635.97\\
				\tiny\scshape z3str3& \tiny197 & \tiny1 & \tiny2 & \tiny0 & \tiny105.07&& \tiny0 & \tiny9 & \tiny0 & \tiny0 & \tiny0.33&& \tiny93 & \tiny43 & \tiny58 & \tiny6 & \tiny2089.92&& \tiny175 & \tiny10 & \tiny12 & \tiny3 & \tiny490.23&& \tiny198 & \tiny1 & \tiny1 & \tiny0 & \tiny41.52\\
				\tiny\scshape z3seq& \tiny183 & \tiny0 & \tiny17 & \tiny0 & \tiny545.24&& \tiny9 & \tiny0 & \tiny0 & \tiny0 & \tiny1.81&& \tiny126 & \tiny0 & \tiny73 & \tiny1 & \tiny2199.99&& \tiny193 & \tiny0 & \tiny6 & \tiny1 & \tiny200.09&& \tiny193 & \tiny0 & \tiny7 & \tiny0 & \tiny217.35\\
				\tiny\scshape norn& \tiny176 & \tiny0 & \tiny20 & \tiny4 & \tiny1037.63&& \tiny0 & \tiny0 & \tiny9 & \tiny0 & \tiny270.00&& \tiny71 & \tiny0 & \tiny128 & \tiny1 & \tiny4038.83&& \tiny60 & \tiny0 & \tiny72 & \tiny68 & \tiny3216.95&& \tiny112 & \tiny0 & \tiny9 & \tiny79 & \tiny742.34\\
				\tiny\scshape sloth& \tiny101 & \tiny0 & \tiny99 & \tiny0 & \tiny3658.34&& \tiny7 & \tiny0 & \tiny2 & \tiny0 & \tiny124.56&& \tiny121 & \tiny0 & \tiny67 & \tiny12 & \tiny2808.48&& \tiny16 & \tiny0 & \tiny184 & \tiny0 & \tiny5615.81&& \tiny9 & \tiny2 & \tiny187 & \tiny2 & \tiny5750.79

		\end{tabular}}
		\caption{Benchmark results (\faCheckCircleO: correct classified, \faStopCircleO: reported unknown, \faPauseCircleO: timed out after 30 seconds, \faTimesCircleO: incorrectly classified, \faClockO: total Time in seconds) }
		\label{tab:results}\vspace*{-0.3cm}
	\end{table}
	\autoref{tab:results} is read as follows:  \faCheckCircleO{} is the count of instances classified as correctly, where \faTimesCircleO{} marks the incorrect classified cases. For instances marked with \faStopCircleO{} the solver returned no answer but terminated before the timeout of 30 seconds was reached, where in \faPauseCircleO{} marked cases the solver was killed after 30 seconds. The row marked by \faClockO{} states the overall solving time. The produced substitutions were checked regarding their correctness afterwards. The classification of \faTimesCircleO{} was done by ad-hoc case inspection whenever not all solvers agreed on a result.  In the cases one solver produced a valid solution, and others did not, we validated the substitutions manually. For the cases where one solver determined an equation is unsatisfied and all others timed out, we treated the unsat result as correct. This means that we only report errors if a solver reports unsat and we know the equation was satisfiable. During our evaluation of track I CVC4 crashed with a null-pointer exception regarding the word equation $dbebgddbecfcbbAadeeaecAgebegee\-caf\-egebdbagddaadbddcaeeebfabfef\-abfacdgAgaabgegagf \dotequal dbebgddbeA\-fcbbAaIegeeAaD\-e\-gagf$, where lowercase symbols are letters and uppercase symbols are variables. Worth mentioning is the reporting of 14 satisfiable benchmarks by the tool Sloth without being able to produce a valid model, while at least two other tools classified them as unsatisfiable. We treated this as an erroneous behaviour.
		 
%
	 The result shows that \textsc{Woorpje}   produces reliable results (0 errors) in competitive time. 
	 It outperforms the competitors in track I,III and IV and sticks relatively tight to the leaders Z3str3, Z3Seq and CVC4 on track V. On track II \woorpje{} trails CVC4 and Z3Str3. The major inefficiency of \woorpje\  is related to  multiple equations with large alphabets and linear length constraints. 

	 It is worth emphasising,  that  the
     benchmarks developed here seem of intrinsic interest, as
     they challenge even established solvers.

%% file: conclusion.tex
	In this paper we present a method for solving word equations by using a SAT-Solver. The method is implemented in our new tool \woorpje\ and experiments show it is competitive with state-of-the-art string solvers. 
\woorpje\ solves word equations instances that other solvers fail to solve. This indicates that our technique can complement existing in a portfolio approach.

	In the future, aim to extend our approach  to include regular constraints. As our approach rely on automata theory, it is expected that this could be achievable. Another step is the enrichment of our linear constraint solving. We currently  do a basic analysis by using the MDDS. There are a few refinement steps described in \cite{abio2014encoding} which seem applicable. A next major step is to develop a more efficient encoding of the alphabet of constants. Currently the state space explodes due to the massive branching caused by the usage of large alphabets.  

%% file: llncs.bbl
\begin{thebibliography}{20}
\providecommand{\natexlab}[1]{#1}
\providecommand{\url}[1]{\texttt{#1}}
\providecommand{\urlprefix}{}

\bibitem[{Abdulla et~al.(2015)Abdulla, Atig, Chen, Hol{\'i}k, Rezine,
  R{\"u}mmer, and Stenman}]{abdulla2015}
Abdulla, P.A., Atig, M.F., Chen, Y.F., Hol{\'i}k, L., Rezine, A., R{\"u}mmer,
  P., Stenman, J.: Norn: An {SMT} solver for string constraints.
\newblock In: Kroening, D., P{\u{a}}s{\u{a}}reanu, C.S. (eds.) Computer Aided
  Verification. pp. 462--469. Springer International Publishing, Cham (2015)

\bibitem[{Ab{\'\i}o and Stuckey(2014)}]{abio2014encoding}
Ab{\'\i}o, I., Stuckey, P.J.: Encoding linear constraints into {SAT}.
\newblock In: International Conference on Principles and Practice of Constraint
  Programming. pp. 75--91. Springer (2014)

\bibitem[{Audemard and Simon(2018)}]{audemard2018glucose}
Audemard, G., Simon, L.: On the glucose {SAT} solver.
\newblock International Journal on Artificial Intelligence Tools 27(01),
  1840001 (2018)

\bibitem[{Barrett et~al.(2011)Barrett, Conway, Deters, Hadarean, Jovanovi{\'c},
  King, Reynolds, and Tinelli}]{barrett2011}
Barrett, C., Conway, C.L., Deters, M., Hadarean, L., Jovanovi{\'c}, D., King,
  T., Reynolds, A., Tinelli, C.: {CVC4}.
\newblock In: International Conference on Computer Aided Verification. pp.
  171--177. Springer (2011)

\bibitem[{Berzish et~al.(2017)Berzish, Ganesh, and Zheng}]{berzish2017}
Berzish, M., Ganesh, V., Zheng, Y.: Z3str3: A string solver with theory-aware
  heuristics.
\newblock In: 2017 Formal Methods in Computer Aided Design (FMCAD). pp. 55--59
  (Oct 2017)

\bibitem[{Bj{\o}rner et~al.(2009)Bj{\o}rner, Tillmann, and
  Voronkov}]{bjorner2009path}
Bj{\o}rner, N., Tillmann, N., Voronkov, A.: Path feasibility analysis for
  string-manipulating programs.
\newblock In: International Conference on Tools and Algorithms for the
  Construction and Analysis of Systems. pp. 307--321. Springer (2009)

\bibitem[{Cadar et~al.(2008)Cadar, Dunbar, and Engler}]{CadarDE08}
Cadar, C., Dunbar, D., Engler, D.R.: {KLEE:} unassisted and automatic
  generation of high-coverage tests for complex systems programs.
\newblock In: Draves, R., van Renesse, R. (eds.) 8th {USENIX} Symposium on
  Operating Systems Design and Implementation, {OSDI} 2008, December 8-10,
  2008, San Diego, California, USA, Proceedings. pp. 209--224. {USENIX}
  Association (2008),
  \urlprefix\url{http://www.usenix.org/events/osdi08/tech/full\_papers/cadar/cadar.pdf}

\bibitem[{Chen et~al.(2019)Chen, Hague, Lin, R{\"u}mmer, and
  Wu}]{chen2019decision}
Chen, T., Hague, M., Lin, A.W., R{\"u}mmer, P., Wu, Z.: Decision procedures for
  path feasibility of string-manipulating programs with complex operations.
\newblock Proceedings of the ACM on Programming Languages 3(POPL), 49 (2019)

\bibitem[{Cordeiro et~al.(2018)Cordeiro, Kesseli, Kroening, Schrammel, and
  Trt{\'{\i}}k}]{jbmc}
Cordeiro, L.C., Kesseli, P., Kroening, D., Schrammel, P., Trt{\'{\i}}k, M.:
  {JBMC:} {A} bounded model checking tool for verifying java bytecode.
\newblock In: Chockler, H., Weissenbacher, G. (eds.) Computer Aided
  Verification - 30th International Conference, {CAV} 2018, Held as Part of the
  Federated Logic Conference, FloC 2018, Oxford, UK, July 14-17, 2018,
  Proceedings, Part {I}. Lecture Notes in Computer Science, vol. 10981, pp.
  183--190. Springer (2018),
  \urlprefix\url{https://doi.org/10.1007/978-3-319-96145-3\_10}

\bibitem[{Day et~al.(2017)Day, Manea, and Nowotka}]{MFCS2017}
Day, J.D., Manea, F., Nowotka, D.: The hardness of solving simple word
  equations.
\newblock In: Proc. {MFCS} 2017. LIPIcs, vol.~83, pp. 18:1--18:14 (2017)

\bibitem[{Hol{\'\i}k et~al.(2017)Hol{\'\i}k, Janků, Lin, R{\"u}mmer, and
  Vojnar}]{holik2017string}
Hol{\'\i}k, L., Janků, P., Lin, A.W., R{\"u}mmer, P., Vojnar, T.: String
  constraints with concatenation and transducers solved efficiently.
\newblock Proceedings of the ACM on Programming Languages 2(POPL), 4 (2017)

\bibitem[{Je{\.z}(2013)}]{jez2013}
Je{\.z}, A.: Recompression: a simple and powerful technique for word equations.
\newblock In: 30th International Symposium on Theoretical Aspects of Computer
  Science, {STACS} 2013, February 27 - March 2, 2013, Kiel, Germany. pp.
  233--244 (2013),
  \urlprefix\url{https://doi.org/10.4230/LIPIcs.STACS.2013.233}

\bibitem[{Je{\.z}(2017)}]{jez2017}
Je{\.z}, A.: Word equations in nondeterministic linear space.
\newblock In: Proc. {ICALP} 2017. LIPIcs, vol.~80, pp. 95:1--95:13. Schloss
  Dagstuhl - Leibniz-Zentrum fuer Informatik (2017)

\bibitem[{Karhum{\"a}ki et~al.(2000)Karhum{\"a}ki, Mignosi, and
  Plandowski}]{karhumaki2000}
Karhum{\"a}ki, J., Mignosi, F., Plandowski, W.: The expressibility of languages
  and relations by word equations.
\newblock Journal of the ACM (JACM) 47(3), 483--505 (2000)

\bibitem[{Kiezun et~al.(2009)Kiezun, Ganesh, Guo, Hooimeijer, and
  Ernst}]{kiezun2009}
Kiezun, A., Ganesh, V., Guo, P.J., Hooimeijer, P., Ernst, M.D.: Hampi: a solver
  for string constraints.
\newblock In: Proceedings of the eighteenth international symposium on Software
  testing and analysis. pp. 105--116. ACM (2009)

\bibitem[{Makanin(1977)}]{makanin1977}
Makanin, G.S.: The problem of solvability of equations in a free semigroup.
\newblock Sbornik: Mathematics 32(2), 129--198 (1977)

\bibitem[{Plandowski and Rytter(1998)}]{PlandowskiR98}
Plandowski, W., Rytter, W.: Application of {L}empel-{Z}iv encodings to the
  solution of words equations.
\newblock In: Proc. 25th International Colloquium on Automata, Languages and
  Programming, ICALP'98. Lecture Notes in Computer Science, vol. 1443, pp.
  731--742. Springer (1998)

\bibitem[{Plandowski(1999)}]{plandowski1999}
Plandowski, W.: Satisfiability of word equations with constants is in pspace.
\newblock In: Foundations of Computer Science, 1999. 40th Annual Symposium on.
  pp. 495--500. IEEE (1999)

\bibitem[{Saxena et~al.(2010)Saxena, Akhawe, Hanna, Mao, McCamant, and
  Song}]{saxena2010symbolic}
Saxena, P., Akhawe, D., Hanna, S., Mao, F., McCamant, S., Song, D.: A symbolic
  execution framework for javascript.
\newblock In: 2010 IEEE Symposium on Security and Privacy. pp. 513--528. IEEE
  (2010)

\bibitem[{Trinh et~al.(2016)Trinh, Chu, and Jaffar}]{trinh2016}
Trinh, M.T., Chu, D.H., Jaffar, J.: Progressive reasoning over
  recursively-defined strings.
\newblock In: Chaudhuri, S., Farzan, A. (eds.) Computer Aided Verification. pp.
  218--240. Springer International Publishing, Cham (2016)

\end{thebibliography}
